\documentclass[onecolumn,prd,amsmath,amssymb,nofootinbib]{revtex4}

\usepackage{graphicx}
\usepackage{dcolumn}
\usepackage{bm}

\begin{document}

\title{Comment on ``Multidimensional arrow of time'' (arXiv:2601.14134)}

\author{Andrei Galiautdinov}
\email{ag1@uga.edu}
\affiliation{
Department of Physics and Astronomy, 
University of Georgia, Athens, Georgia 30602, USA
}

\date{\today}

\begin{abstract}
In a recent preprint [arXiv:2601.14134v1], Rubin argues 
that the arrow of time originates from the monotonic 
growth of the volume of extra dimensions. While the 
identification of a geometric origin for time's arrow is 
compelling in the case of brane-world models, we point 
out a possible tension between the proposed volume 
growth and the observational stability of the effective 
four-dimensional Newton's gravitational constant, 
$G_N$, that may arise in Kaluza-Klein (KK) theory. 
In standard KK approaches, such volume 
growth induces a time-variation of $G_N$ that exceeds 
Big Bang Nucleosynthesis (BBN) and Lunar Laser Ranging 
(LLR) bounds by many orders of magnitude. To resolve 
this tension while preserving the author's key insight 
in the Kaluza-Klein case, we propose an extension:
the ``shape-dynamic arrow of time.'' By utilizing the 
scale-invariant monotonicity of Perelman's $\nu$-entropy 
under normalized Ricci flow, we demonstrate how 
an arrow of time can emerge from the geometric 
smoothing of extra dimensions at fixed volume, thereby 
satisfying observational constraints on fundamental constants.
\end{abstract}

\maketitle

\section{Introduction}

The origin of the arrow of time remains one of the premier 
open problems in physics. In Ref.~\cite{Rubin2026}, the 
author proposes a novel geometric mechanism, the 
Multidimensional Arrow of Time (MAT). The central 
hypothesis is that the universe possesses extra spatial 
dimensions whose volume, $V_{int}$, evolves monotonically. 
This volume growth serves as a master clock, defining 
the direction of time independently of the thermodynamic 
entropy of matter fields.

While the MAT proposal offers an elegant escape from 
the thermodynamic definition of time, it faces  
tension with Kaluza-Klein (KK) phenomenology. In this 
Comment, we first quantify the variations in fundamental 
constants implied by the MAT mechanism in the KK case, 
demonstrating that they appear to conflict with current observational 
bounds (Section \ref{sec:stability}). In Section \ref{sec:extension}, 
we propose a constructive extension: replacing the \textit{volume} 
growth with \textit{shape} evolution 
({\it cf}.\ \cite{Barbour2014a,Barbour2014}). 
By invoking Perelman's results on the entropy of Ricci 
flow \cite{Perelman2002}, we define a geometric arrow 
of time that preserves fundamental physical constants.

\section{The Stability of Newton's Constant}
\label{sec:stability}

The standard action for gravity in a $(4+n)$-dimensional 
spacetime is given by \cite{Rubin2026},
\begin{equation}
S = \frac{1}{16\pi G_{D}} \int d^4x \, d^n y \sqrt{-g^{(D)}} \, R^{(D)},
\end{equation}
where $G_D$ is the fundamental $D$-dimensional 
gravitational constant ($D=4+n$), and the coordinates 
are decomposed as $X^M = (x^\mu, y^a)$. Following 
\cite{Rubin2026}, we assume a product metric \emph{Ansatz},
\begin{equation}
ds^2 
= 
g_{\mu\nu}^{(4)}(x)dx^\mu dx^\nu 
- b^2(t) \gamma_{ab}(y)dy^a dy^b,
\end{equation}
where $\gamma_{ab}$ is the metric of the internal 
manifold $\mathcal{K}$ and $b(t)$ is the scale factor 
(modulus field) of the internal space. The effective 
four-dimensional action is obtained by integrating over 
the internal manifold,
\begin{equation}
S_{eff} 
= 
\frac{V_n(t)}{16\pi G_{D}} \int d^4x \sqrt{-g^{(4)}} \, R^{(4)} 
+ \dots ,
\end{equation}
where the physical volume of the internal space is 
$V_{n}(t) = \int d^n y \sqrt{\gamma} \, b(t)^n \equiv V_0 \, b(t)^n$.
The effective 4D Planck mass squared, 
$M_{Pl}^2 = (8\pi G_N)^{-1}$, is directly proportional to 
this volume, or,
\begin{equation} 
\label{eq:G_relation}
G_N(t) = \frac{G_D}{V_{n}(t)}. 
\end{equation}
Consequently, any time evolution of the extra-dimensional 
volume induces a variation in Newton's constant,
\begin{equation}
\frac{\dot{G}_N}{G_N} 
= 
-\frac{\dot{V}_{n}}{V_{n}} 
= 
- n \frac{\dot{b}}{b}.
\label{eq:G_variation}
\end{equation}

Current observational limits on the variation of $G_N$ 
are stringent. Lunar Laser Ranging (LLR) experiments 
constrain the relative variation to \cite{Williams2004},
\begin{equation}
\label{eq:G_variation_numeric}
\left| \frac{\dot{G}_N}{G_N} \right| \lesssim 10^{-13} \, \text{yr}^{-1}.
\end{equation}
Furthermore, constraints from Big Bang Nucleosynthesis 
(BBN) imply that the value of $G_N$ at redshift $z \sim 10^9$ 
could not have differed from the current value by more than 
a few percent \cite{Uzan2011}.

The MAT hypothesis \cite{Rubin2026} relies on 
$\dot{V}_{n} > 0$ to define the arrow of time. If this growth 
is the primary driver of macroscopic time directionality in 
Kaluza-Klein theory, it implies a significant rate of change. 
If $\dot{V}_{n}/V_{n}$ is large enough to distinguish 
``past'' from ``future'' on relevant timescales, it violates Eq.~\eqref{eq:G_variation_numeric}. 
Conversely, if the growth is suppressed to satisfy LLR 
bounds ($\sim 10^{-13}$), the arrow becomes virtually 
static, raising the question of how such a minuscule drift 
effectively suppresses backward fluctuations in macroscopic 
processes. 

\section{Proposed Extension (The Shape-Dynamic Arrow)}
\label{sec:extension}

\subsection{The main idea}
\label{sec:mainIdea}

To rescue the geometric arrow of time from the constraints 
of constant stability in the KK case, we propose a modification: 
the geometry of the extra dimensions evolves, but its volume 
remains fixed. This can be realized using the framework 
of \textit{Normalized Ricci Flow}.\footnote{We assume the 
standard Kaluza-Klein ansatz where the total spacetime 
manifold $\mathcal{M}_{total}$ is locally a product 
$\mathcal{M}_4 \times \mathcal{K}$. 
While $\mathcal{M}_{total}$ is pseudo-Riemannian (Lorentzian), the 
internal manifold of extra dimensions, $\mathcal{K}$, is compact 
and Riemannian (strictly spatial). Furthermore, we assume the internal 
dimensionality $n \ge 2$, as one-dimensional manifolds are flat ($R=0$) 
and thus possess no intrinsic shape dynamics to drive the flow.} 
In his seminal work on the Geometrization Conjecture \cite{Perelman2002}, 
Perelman introduced a statistical mechanical framework for geometric evolution. 
He defined the entropy functional, denoted $\mathcal{W}$, 
which acts on the metric $g_{ij}$ of the $n$-dimensional 
internal manifold, a smooth scalar potential function $f$, 
and a strictly positive scale parameter $\tau$. The functional is 
defined as
\begin{equation}
\mathcal{W}(g_{ij}, f, \tau) 
= 
\int_{\mathcal{K}} 
\left[ \tau \left( R + |\nabla f|^2 \right) + f - n \right] u \, dV,
\label{eq:W_functional}
\end{equation}
where $dV = \sqrt{\det g} \, d^n y$ is the standard volume 
element, $R$ is the scalar curvature (representing the local 
geometric complexity), $f$ is the dilaton potential, $u$ is the conjugate heat kernel 
(defined as $u = (4\pi \tau)^{-n/2} e^{-f}$, satisfying $\int_{\mathcal{K}} u \, dV = 1$), 
and $\tau$ is the scale parameter (evolving as $\dot{\tau} = -1$). 
If the metric evolves according to the Ricci flow, 
\begin{equation}
\partial_t g_{ij} = -2R_{ij}, 
\end{equation}
and the scalar function $f$ evolves according to a conjugate 
heat equation, Perelman proved the monotonicity formula,
\begin{equation}
\frac{d}{dt} \mathcal{W}(g(t), f(t), \tau(t)) 
= 
2 \int_{\mathcal{K}} \tau \left| R_{ij} 
+ \nabla_i \nabla_j f - \frac{1}{2\tau} g_{ij} \right|^2 u \, dV \ge 0.
\label{eq:monotonicity}
\end{equation}
Eq.\ \eqref{eq:monotonicity} is the geometric equivalent 
of Boltzmann's H-theorem. It says that the ``geometric 
entropy'' $\mathcal{W}$ is strictly non-decreasing along the 
flow. The system evolves irreversibly from high-complexity, 
irregular shapes (lower entropy) toward constant-curvature 
or soliton configurations (maximum entropy). 
Our proposal is that the internal manifold $\mathcal{K}$ evolves 
according to the \textit{volume-normalized} Ricci flow equation,
\begin{equation}
\label{eq:norm_flow}
\frac{\partial \gamma_{ab}}{\partial t} 
= -2 R_{ab} + \frac{2}{n} \langle R \rangle \gamma_{ab},
\end{equation}
with $\langle R \rangle = V_n^{-1} \int R \sqrt{\gamma} d^n y$ 
being the average scalar curvature, in which case the key role 
is played by Perelman's scale-invariant $\nu$-entropy
(see Subsection \ref{app:monotonicity} below and Appendix \ref{app:proof} for additional technical details; also, {\it cf}.\ \cite{Headrick2006}). 
The physical justification for this scenario may be given by deriving 
the Ricci flow equation directly from the full $D$-dimensional 
Einstein Field Equations (and then imposing the 
volume-conservation constraint). While the full theory 
is Lorentzian and second-order (hyperbolic), let us see how 
the Ricci flow may emerge as the effective dynamics 
of the internal Riemannian metric in the high-friction 
(overdamped) limit. 

\subsection{Emergence of the Ricci flow in overdamped limit}

Consider a $D$-dimensional spacetime with coordinates 
$X^M = (t, x^i, y^a)$, and adopt the metric ansatz,
\begin{equation}
ds^2 
= 
-dt^2 + a^2(t) \delta_{ij} dx^i dx^j + \gamma_{ab}(t, y) dy^a dy^b.
\end{equation}
The internal components of the vacuum Einstein equations, $R_{ab}^{(D)} = 0$, 
yield a wave equation for the internal geometry,
\begin{equation}
\ddot{\gamma}_{ab} 
+ \left( 3H + \frac{1}{2} \text{tr}(\gamma^{-1}\dot{\gamma}) \right) \dot{\gamma}_{ab} 
- 2 (\text{nonlinear terms}) = -2 R_{ab}^{(\gamma)}.
\label{eq:wave_eqn}
\end{equation}
The term in the parentheses acts as a friction coefficient $\zeta \approx 3H$ 
(assuming volume stabilization $\dot{V}_n \approx 0$).

We next apply the \emph{Overdamped Approximation}. 
We assume the early universe is in a high-friction regime 
where the Hubble damping $3H$ is large compared to 
the inertial acceleration of the internal shape $\ddot{\gamma}_{ab}$.
Under the condition $|\ddot{\gamma}_{ab}| \ll | 3H \dot{\gamma}_{ab} |$, 
Eq.~(\ref{eq:wave_eqn}) reduces to a first-order system,
\begin{equation}
3H \dot{\gamma}_{ab} \approx -2 R_{ab}^{(\gamma)}.
\end{equation}
Absorbing the positive scalar factor $(3H)^{-1}$ into a redefinition 
of the time coordinate $d\tau = (3H)^{-1} dt$, we recover the Ricci flow equation,
\begin{equation}
\frac{\partial \gamma_{ab}}{\partial \tau} = -2 R_{ab}^{(\gamma)}.
\end{equation}
Adding the volume constraint (Appendix \ref{app:physics}) restores 
the normalization term.\footnote{The validity of the 
overdamped approximation requires $H \gg L_{int}^{-1}$ 
(where $L_{int}$ is the scale of extra dimensions). 
This condition is satisfied during Inflation but violated in 
the current epoch. 
Consequently, the ``Shape-Dynamic Arrow'' is primarily a \textit{primordial} mechanism. 
It drives the universe from a generic initial state to a maximum-entropy (constant curvature) 
configuration in the very early universe. In the late universe, the shape degrees of freedom 
either freeze into this soliton configuration or oscillate as massive Kaluza-Klein modes, 
effectively stopping the geometric clock. Thus, this mechanism explains the \textit{origin} 
of the arrow of time (the smooth initial boundary condition) rather than its current propagation.}

\subsection{Monotonicity under normalized flow}
\label{app:monotonicity}

In Subsection \ref{sec:mainIdea}, we invoked the monotonicity 
of Perelman's entropy to define the arrow of time. However, 
Perelman's original derivation \cite{Perelman2002} applies to 
the unnormalized Ricci flow,
\begin{equation}
\partial_t g_{ij} = -2R_{ij},
\end{equation} 
which shrinks the manifold volume. Let us then verify that 
a monotonic quantity also exists for the volume-normalized 
flow used in our proposal. We see that as follows:

Perelman defined the entropy $\mathcal{W}(g, f, \tau)$ depending 
on a scale parameter $\tau$. To construct a quantity that depends 
only on the geometry (independent of scale and the auxiliary 
function $f$), he introduced the $\nu$-entropy.
First, minimize $\mathcal{W}$ over all probability densities 
(determined by $f$),
\begin{equation}
\mu(g, \tau) 
= 
\inf_{f} \mathcal{W}(g, f, \tau), \quad \text{subject to } 
\int (4\pi\tau)^{-n/2} e^{-f} dV = 1.
\end{equation}
Next, minimize over the scale parameter $\tau$,
\begin{equation}
\nu(g) = \inf_{\tau > 0} \mu(g, \tau).
\end{equation}
The quantity $\nu(g)$ is a strictly geometric invariant which 
is \emph{scale invariance}. Namely, for any positive constant 
$c > 0$,
\begin{equation}
\nu(c \cdot g) = \nu(g).
\label{eq:scale_invariance}
\end{equation}
This reflects the fact that $\nu$ measures the complexity 
of the \textit{shape}, not the volume.

Let $g(t)$ be a solution to the standard Ricci flow, and let 
$\bar{g}(\bar{t})$ be a solution to the volume-normalized 
Ricci flow defined in Eq.\ (\ref{eq:norm_flow}).
It is a standard result in geometric analysis \cite{Hamilton1982} 
that the normalized flow is equivalent to the unnormalized flow 
up to a time-dependent rescaling (homothety) and a 
reparameterization of time. Specifically, the metrics are related by
\begin{equation}
\bar{g}(\bar{t}) = \psi(t) g(t),
\end{equation}
where $\psi(t)$ is a scalar function that rescales the volume to 
be constant. We then have the following theorem.

\textbf{Theorem:} The $\nu$-entropy is monotonically 
non-decreasing under volume-normalized 
Ricci flow.

\textbf{Proof:} 
Perelman proved that $\nu(g(t))$ is strictly increasing 
under standard Ricci flow (unless the manifold is a Ricci soliton):
\begin{equation}
\frac{d}{dt} \nu(g(t)) \ge 0.
\end{equation}
By the scale invariance property (Eq.~\ref{eq:scale_invariance}), 
the entropy of the normalized metric is identical to the entropy 
of the unnormalized metric at the corresponding unnormalized time,
\begin{equation}
\nu(\bar{g}(\bar{t})) = \nu(\psi(t) g(t)) = \nu(g(t)).
\end{equation}
Since the time reparameterization $\frac{d\bar{t}}{dt} > 0$ 
preserves the direction of time, the monotonicity carries over,
\begin{equation}
\frac{d}{d\bar{t}} \nu(\bar{g}) 
= 
\left( \frac{dt}{d\bar{t}} \right) \frac{d}{dt} \nu(g) \ge 0.
\end{equation}
Thus, the ``shape-dynamic arrow of time'' is mathematically 
well-defined for the fixed-volume evolution. Rubin's Postulate \cite{Rubin2026}, 
according to which the grows of time is the consequence of 
the entropy grows, then fixes the arrow's direction.
\footnote{The motivation comes from the earlier important 
proposals by Penrose, Cort\^{e}s, Smolin, and others (see, e.\ g., \cite{Penrose1979,Cortes2014,Cortes2018})} As a result, the universe evolves monotonically toward a state of maximum $\nu$-entropy
while satisfying observational constraints on fundamental constants.

\section{Conclusion}
\label{sec:conclusion}

The proposal in \cite{Rubin2026} identifies a profound link 
between extra-dimensional geometry and the arrow of time. 
While the original formulation relying on volume growth faces 
tension with precision tests of gravity in Kaluza-Klein theory
($|\dot{G}/G|$ constraints) , 
we have outlined a natural extension using Perelman's entropy and 
normalized Ricci flow. By shifting the definition of the arrow from the 
\textit{modulus} (volume) to the \textit{shape} (conformal geometry), 
one recovers a monotonic clock that is decoupled from local coupling 
constants but remains fundamental to the topology of the bulk.
As a result, our approach preserves the essential spirit of Rubin's original 
proposal, while effectively establishing the ``geometric initial conditions''---the 
low-entropy boundary term---required for the subsequent thermodynamic and cosmological 
arrows of time, thus ensuring the stability of physical laws in the current epoch.

\appendix

\section{Proof of Volume Conservation}
\label{app:proof}

In this Appendix, we demonstrate that the volume-normalized 
Ricci flow strictly preserves the total volume of the internal 
manifold $\mathcal{K}$, thereby ensuring the stability of the 
effective four-dimensional Newton's gravitational constant $G_N$.

Let $\mathcal{K}$ be a compact $n$-dimensional Riemannian 
manifold with metric $g_{ij}(t)$. The total volume is defined as
\begin{equation}
V_n(t) = \int_{\mathcal{K}} \sqrt{g} \, d^n y,
\end{equation}
where $\sqrt{g} = \sqrt{\det(g_{ij})}$ is the volume density.
The rate of change of the volume is found by differentiating,
\begin{equation}
\frac{dV_n}{dt} 
= \int_{\mathcal{K}} \frac{\partial}{\partial t} 
\left( \sqrt{g} \right) d^n y.
\end{equation}
Using the standard identity for the variation of the determinant, 
$\frac{\partial}{\partial t} \sqrt{g} = \frac{1}{2} \sqrt{g} g^{ij} \dot{g}_{ij}$, 
we substitute the flow equation proposed in Eq.\ (\ref{eq:norm_flow}),
\begin{equation}
\dot{g}_{ij} = -2 R_{ij} + \frac{2}{n} \langle R \rangle g_{ij}.
\end{equation}
Taking the trace of this equation gives
\begin{equation}
g^{ij} \dot{g}_{ij} = -2 g^{ij} R_{ij} + \frac{2}{n} \langle R \rangle g^{ij} g_{ij}.
\end{equation}
Recalling that $g^{ij} R_{ij} = R$ (the scalar curvature) 
and $g^{ij} g_{ij} = n$, this simplifies to
\begin{equation}
g^{ij} \dot{g}_{ij} = -2 R + 2 \langle R \rangle.
\end{equation}
Substituting this back into the volume derivative, we get,
\begin{equation}
\frac{dV_n}{dt} 
= \int_{\mathcal{K}} \frac{1}{2} 
\left( -2 R + 2 \langle R \rangle \right) \sqrt{g} \, d^n y 
= \int_{\mathcal{K}} (\langle R \rangle - R) \sqrt{g} \, d^n y.
\end{equation}
Since $\langle R \rangle$ is spatially constant, it factors out 
of the integral,
\begin{equation}
\frac{dV_n}{dt} 
= 
\langle R \rangle \int_{\mathcal{K}} \sqrt{g} \, d^n y 
- \int_{\mathcal{K}} R \sqrt{g} \, d^n y.
\end{equation}
By definition, the first integral is the volume $V_n$, and 
$\langle R \rangle \equiv V_n^{-1} \int R \sqrt{g} d^n y$. 
Therefore,
\begin{equation}
\frac{dV_n}{dt} 
= 
\left( \frac{1}{V_n} \int_{\mathcal{K}} R \sqrt{g} \, d^n y \right) V_n 
- \int_{\mathcal{K}} R \sqrt{g} \, d^n y = 0.
\end{equation}
Thus, $V_n(t)$ is constant in time. By the Kaluza-Klein relation 
$G_N \propto V_n^{-1}$, it follows that $\dot{G}_N = 0$.

\section{Volume Constrained Dynamics}
\label{app:physics}

As we pointed out in Section \ref{sec:stability}, the volume of the internal 
space must remain fixed to satisfy observational constraints on $G_N$. 
However, standard gradient flow (unnormalized Ricci flow) naturally seeks 
to shrink the manifold (for positive curvature). To address this 
formally, we introduce a time-dependent Lagrange multiplier 
$\lambda(t)$ to enforce the constraint 
$\int \sqrt{\gamma} d^n y = V_0$ and modify the standard flow equation as
\begin{equation}
\frac{\partial \gamma_{ab}}{\partial t} 
= 
-2 R_{ab} + \lambda(t) \gamma_{ab}.
\end{equation}
To determine $\lambda(t)$, we impose the condition 
$\frac{d}{dt} V_n = 0$. 
Using the identity $\partial_t \sqrt{\gamma} 
= \frac{1}{2}\sqrt{\gamma} \gamma^{ab} \dot{\gamma}_{ab}$, 
we then take the trace of the flow equation,
\begin{equation}
\gamma^{ab} \dot{\gamma}_{ab} = -2R + n \lambda(t).
\end{equation}
Integrating the volume rate of change over the manifold, we get
\begin{equation}
\frac{dV_n}{dt} = \int_{\mathcal{K}} \frac{1}{2} (-2R + n\lambda) \sqrt{\gamma} \, d^n y = 0,
\end{equation}
which simplifies to
\begin{equation}
-\int_{\mathcal{K}} R \sqrt{\gamma} \, d^n y 
+ \frac{n \lambda}{2} \int_{\mathcal{K}} \sqrt{\gamma} \, d^n y 
= 0.
\end{equation}
Identifying the volume $V_n$ and the average curvature 
$\langle R \rangle V_n = \int R \sqrt{\gamma} d^n y$, 
we solve for $\lambda$,
\begin{equation}
\lambda(t) = \frac{2}{n} \langle R \rangle.
\end{equation}
Substituting this back into the flow equation results in the 
volume-normalized Ricci flow,
\begin{equation}
\frac{\partial \gamma_{ab}}{\partial t} 
= 
-2 R_{ab} + \frac{2}{n} \langle R \rangle \gamma_{ab}.
\end{equation}
This confirms that the normalized flow is the unique geometric evolution 
that maximizes entropy while rigidly preserving the scale of the extra dimensions.

\section{Physical Interpretation of the Measure $u$}
\label{app:measure}

Perelman's monotonicity theorem relies on a scalar density 
$u = (4\pi\tau)^{-n/2} e^{-f}$ evolving via the conjugate heat equation,
\begin{equation}
\square^* u = \left( -\frac{\partial}{\partial t} - \Delta + R \right) u = 0.
\end{equation}
In the absence of matter fields, the physical status of $u$ requires 
clarification. We interpret $u$ not as a matter distribution, but as 
the \emph{spectral density} of the vacuum geometry itself.

In quantum field theory on curved spacetime with extra 
dimensions the vacuum structure is determined by the 
spectral properties of the Laplacian operator 
$\Delta_g$ on the internal manifold. The function $u(y, \tau)$ 
corresponds to the diagonal of the \textit{Heat Kernel} $K(y, y', \tau)$, 
which describes the propagation of vacuum fluctuations or gravitational 
waves across the manifold.

The conjugate heat equation is the condition that preserves the 
spectral definition of the entropy as the background geometry 
evolves. Specifically, Perelman's entropy $\mathcal{W}$ can be 
understood as the Shannon entropy of the probability distribution 
of these spectral vacuum fluctuations.
Thus, the "arrow of time" may be viewed as the irreversible 
loss of information regarding the initial spectral asymmetry of 
the extra dimensions. As the geometry flows toward a constant 
curvature soliton, the spectrum of the Laplacian simplifies, and 
the geometric entropy $\mathcal{W}$ maximizes. This process 
is intrinsic to the spacetime fabric and proceeds without any 
reference to external matter fluids.

\begin{acknowledgments}
The author thanks Sergey Rubin for stimulating correspondence.
\end{acknowledgments}

\end{document}